# COVID-19 Classification of X-ray Images Using Deep Neural Networks

Elisha Goldstein[1]*, Daphna Keidar[2]*, Daniel Yaron[3]*, Yair Shachar[4], Ayelet Blass, Leonid Charbinsky MD[5], Israel Aharony MD[5], Liza Lifshitz MD[5], Dimitri Lumelsky MD[5], Ziv Neeman MD[5], Matti Mizrachi MD[6], Majd Hajouj MD[6], Nethanel Eizenbach MD[6], Eyal Sela MD[6], Chedva S Weiss MD[7], Philip Levin MD[7], Ofer Benjaminov MD[7], Gil N Bachar MD[8], Shlomit Tamir MD[8], Yael Rapson MD[8], Dror Suhami MD[8], Amiel A Dror MD PhD[6], Naama R Bogot MD[7], Ahuva Grubstein MD[8], Nogah Shabshin MD[5], Yishai M Elyada PhD[9], Yonina C Eldar PhD[3]

* Equal contribution

**Abbreviations**

CXR - chest x-ray

COVID-19 - Coronavirus Disease 2019

RT-PCR - reverse transcription polymerase chain reaction

ROC - receiver operating characteristic

P-R curve - precision-recall curve

AUC - area under the curve

[1] Bioinformatics Unit, Life Sciences Core Facilities, Weizmann Institute of Science, Rehovot, Israel
[2] ETH Zürich, D-infk, Rämistrasse 101 8092 Zürich
3 Dept. of Math and Computer Science, Weizmann Institute of Science, Rehovot, Israel
[4] Eyeway Vision Ltd., Yoni Netanyahu St 3, Or Yehuda
[5] Department of Radiology, HaEmek Medical Center, Afula, Israel
[6] Department of Otolaryngology, Head and Neck Surgery, Galilee Medical Center, Nahariya, Israel; The Azrieli Faculty of Medicine, Bar-Ilan University, Safed, Israel
[7] Cardiothoracic Imaging Unit, Shaare Zedek Medical Center, Jerusalem, Israel
[8] Radiology department, Rabin Medical Center, Jabotinsky Rd 39, Petah Tikva; Sakler School of Medicin, Tel-Aviv University, Ramat Aviv, Tel-Aviv
[9] Mobileye Vision Technologies, Ltd., Hartom 13, Jerusalem

GT - ground truth

FPR - false positive rate

TPR - true positive rate

# Abstract

**Background**

In the midst of the coronavirus disease 2019 (COVID-19) outbreak, chest X-ray (CXR) imaging is playing an important role in the diagnosis and monitoring of patients with COVID-19. Machine learning solutions have been shown to be useful for X-ray analysis and classification in a range of medical contexts.

**Purpose**

The purpose of this study is to create and evaluate a machine learning model for diagnosis of COVID-19, and to provide a tool for searching for similar patients according to their X-ray scans.

**Materials and Methods**

In this retrospective study, a classifier was built using a pre-trained deep learning model (ReNet50) and enhanced by data augmentation and lung segmentation to detect COVID-19 in frontal CXR images collected between January 2018 and July 2020 in four hospitals in Israel. A nearest-neighbors algorithm was implemented based on the network results that identifies the images most similar to a given image. The model was evaluated using accuracy, sensitivity, area under the curve (AUC) of receiver operating characteristic (ROC) curve and of the precision-recall (P-R) curve.

**Results**

The dataset sourced for this study includes 2362 CXRs, balanced for positive and negative COVID-19, from 1384 patients (63 +/- 18 years, 552 men). Our model achieved 89.7% (314/350) accuracy and 87.1% (156/179) sensitivity in classification of COVID-19 on a test dataset comprising 15% (350 of 2326) of the original data, with AUC of ROC 0.95 and AUC of the P-R curve 0.94. For each image we retrieve images with the most similar DNN-based image embeddings; these can be used to compare with previous cases.

**Conclusion**

Deep Neural Networks can be used to reliably classify CXR images as COVID-19 positive or negative. Moreover, the image embeddings learned by the network can be used to retrieve images with similar lung findings.

# Summary

Deep Neural Networks and can be used to reliably predict chest X-ray images as positive for coronavirus disease 2019 (COVID-19) or as negative for COVID-19.

# Key Results

- A machine learning model was able to detect chest X-ray (CXR) images of patients tested positive for coronavirus disease 2019 with accuracy of 89.7%, sensitivity of 87.1% and area under receiver operating characteristic curve of 0.95.
- A tool was created for finding existing CXR images with imaging characteristics most similar to a given CXR, according to the model's image embeddings.

# 1 Introduction

The Coronavirus Disease 2019 (COVID-19) pandemic, caused by the SARS-CoV-2 virus, poses tremendous challenges to healthcare systems around the world, and requires physicians to make clinical decisions with limited prior knowledge. Medical decisions are based also on imaging, and can be supported by a method for automatically retrieving prior patients that had similar imaging findings. Moreover, an ongoing concern is to rapidly identify and isolate SARS-CoV-2 carriers in order to contain the disease.

The prevalent test used for COVID-19 identification is Reverse Transcription Polymerase Chain Reaction (RT-PCR) (1,2). However, a recent study suggests that RT-PCR tests result in up to 30% false negatives, depending on the respiratory specimens (3), possibly from non-specific amplification and sample contamination. Taken together, the prominent undetected fraction of active patients inevitably leads to uncontrolled viral dissemination, masking hidden essential epidemiological data (4–6). Additionally, RT-PCR testing kits are expensive and processing them requires dedicated personnel and can take days. Characteristics of COVID-19 such as consolidations and ground-glass opacities can be identified in both CXRs and CT scans (5,7,8). Both are often used to support RT-PCR diagnosis, and are strong candidates for alternative means of COVID-19 testing.

Portable X-ray machines play a central role in COVID-19 handling (9), and most available CXRs of patients with COVID-19 in Israel come from portable X-rays. While COVID-19 is easier to detect in CT (10), CT is more expensive, exposes the patient to higher radiation, and its decontamination process is lengthy and causes severe delays between patients.

Deep learning models have shown impressive abilities in image related tasks, including in many radiological contexts (11,12). They have great potential in assisting COVID-19 management efforts, but require large amounts of training data. When training neural networks for image classification, images from different classes should only differ in the task specific characteristics; it is important, therefore, that all images are taken from the same machines. Otherwise, the network could learn the differences, e.g., between machines associated with different classes rather than identifying physiological and anatomical COVID-19 characteristics.

This study aims to provide machine learning tools for COVID-19 identification and management. A large dataset of images from portable X-rays was sourced and used to train a network that can detect COVID-19 in the images with high reliability and to develop a tool for retrieving CXR images that are similar to each other. The network affords a detection accuracy of 89.7% and sensitivity of 87.1%.

# 2 Materials and Methods

## Approval statement

This retrospective study was approved by the Institutional Review Board (IRB) and the Helsinki committee of the participating medical centers in compliance with the public health regulations and provisions of the current harmonized international guidelines for good clinical practice (ICH-GCP) and in accordance with Helsinki principles. Informed consent was waived by the IRB for the purpose of this study.

## Data and patients

The code development and analysis was performed by six of the authors who are not radiologists (Y.E., D.K., D.Y., Y.S., E.G., A.B.). The clinical images were collected and approved by the authors (L.C., E.A., L.L, D.L., Z.N., M.M., M.H., N.E., E.S., B.N.G, S.T., Y.R., D.S., A.D., N.R.B., A.G., N.S.), who are employed as physicians of multiple disciplines including radiologists in the hospitals which provided the data.

This study includes CXR images from 1384 patients, 360 with a positive COVID-19 diagnosis and 1024 negative, totaling 2427 CXRs. Patients' COVID-19 labels were determined by a combination of RT-PCR testing and clinical assessment by the physicians. The COVID-19 positive images include all CXRs performed with portable X-ray machines on patients admitted to four hospitals in Israel during the pandemic's first wave (December 2019 through April 2020). For the control dataset we obtained CXRs taken by the same X-ray machines prior to December 2019. These are patients without COVID-19, typically with another respiratory disease.

The test set was taken from the full CXR dataset and contains 350 CXR (15%) of which 179 (51%) are positive for COVID-19 and 171 (49%) are negative. To prevent the model from identifying patient-specific image features (e.g., medical implants) and associating them with the label, each patient was either used for the training or the test set.

All images were used in the highest available resolution without lossy compression (e.g. jpeg);

4% (101/2426) of the images were excluded due to lateral positioning, or due to rectangular artifacts in the image, of these 98 were COVID-19 positive. No additional selection criteria were used to exclude CXR images based on clinical radiological findings.

## Image Processing

The model pipeline (Figure 1), begins with a series of preprocessing steps, including augmentation, normalization, and segmentation of the images.

Augmentations are transformations that change features such as image orientation and brightness. These properties are irrelevant for correct classification, but may vary during image acquisition, and can affect the training performance of the network because of its rigid registration with respect to orientation and pixel values. They serve to enlarge the dataset by creating a diverse set of images, increasing model robustness and generalizability (13,14). Importantly, augmentations should correspond to normal variation in CXR acquisition; to ensure this we consulted with radiologists when defining the augmentation parameters (see Appendix).

The normalization process aims to standardize image properties and scale. It consists of cropping black edges, standardizing the brightness and scaling the size of each image to 1024X1024 pixels using bilinear interpolation.

To enhance performance we created an additional image channel using lung segmentation via a U-net (15) pre-trained on a different dataset. This network produces a pixel-mask of the CXR indicating the probability that each pixel belongs in the lungs, allowing the network to access this information while training. Input images contain 3 channels: the original CXR, the segmentation map, and one filled with zeroes. This is done to accommodate the pre-trained models we used that use 3-channel RGB images.

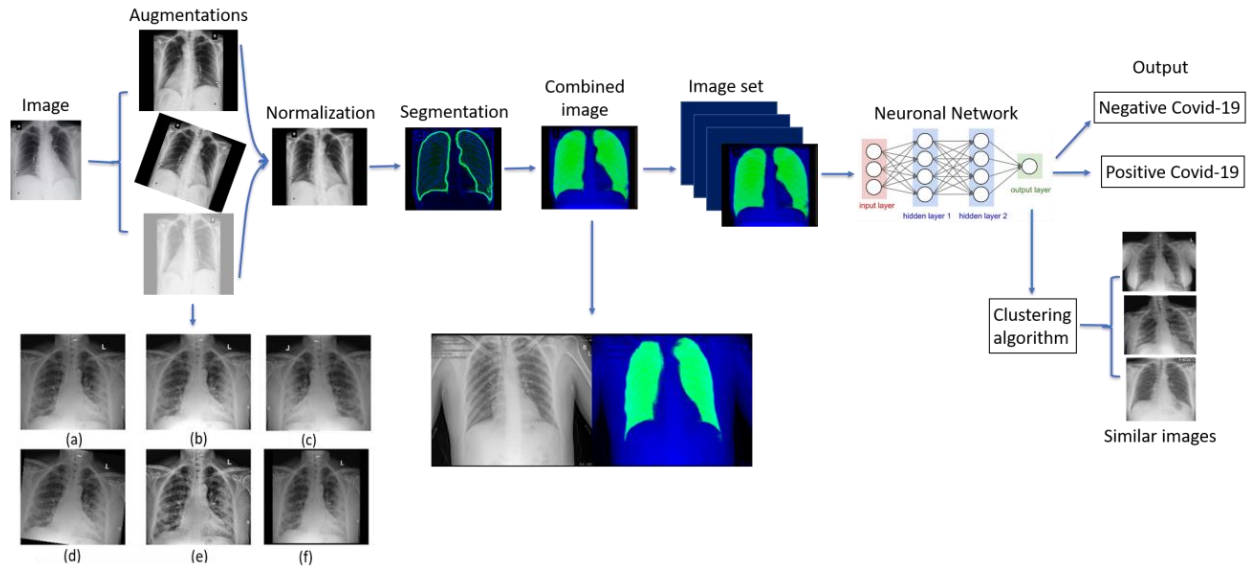

**Figure 1**: Full pipeline workflow overview. First each image undergoes processing consisting of: augmentation, which is a set of visual transformations (transformations shown: (a) original image, (b) brighten, (c) horizontal flip, (d) 7 degrees rotation, (e) CLAHE transformation, (f) scaling), normalization, in order to set a standard scale of image size and color, and segmentation, which emphasizes the area of the lungs and is combined to the image. The entire image set is then fed into a Neuronal Network which produces a classification outcome for each image as positive for coronavirus disease 2019 (COVID-19) or negative for COVID-19. In addition, embedded features are extracted from the last layer of the network and are used to find images with similar characteristics to a given image as learned by the network.

## Network architecture and output

We compared five network models: ResNet34, ResNet50, ResNet152 (16), VGG16 (17) and Chexpert (11). We additionally classify the images by aggregating the results of these networks using a majority vote. The general approach of these architectures is to reduce images from a high-dimensional to a low-dimensional space such that a simple boundary can be used separate image classes. The models were trained using transfer learning, i.e. using pre-trained weights and subsequently retraining them on our data.

Training was performed with the Adam optimizer with an initial learning rate of 1e-6 which was exponentially decreased as epochs progressed. We used cross-entropy as a loss function with an L2 regulariser with regularization coefficient 1e-2. The best test accuracy scores were achieved after 32 epochs. The models were built and trained using Pytorch 1.6; All code will be made available upon publication.

In addition to classification, we propose a method for retrieving a number of CXR images that are the most similar to a given image. The activation of layers of the neural network serve as embeddings of the images into a vector space, and should capture information about clinical indications observed in the images. We use these embeddings to search for similarity between the resulting vectors, and retrieve the nearest neighbors of each image.

## Statistical analysis

For model evaluation we used accuracy, precision, and area under the curve (AUC) for receiver operating characteristic (ROC) and precision recall (P-R) curves.

# 3 Results

## Data acquisition

The patient data included in this study are shown in Table 1. The imaging dataset consists of a total of 2426 CXRs, of which 53% (1289/2426) are positive for COVID-19 and 47% (1138/2426) are negative; 4% (101 of 2426) of the images were excluded due to lateral positioning or having rectangular artifacts covering parts of the image. 98 of these were COVID-19 positive. To our knowledge this is one of the largest datasets of original COVID-19 labeled X-ray images. The demographic statistics of the patients in this study can be seen in Table 1.

**Table 1: Demographic statistics on patients and chest images in this study.**

| Data collection | No. of patients | No. of images** | Sex (men/women/unknown) | Age* |
|---|---|---|---|---|
| COVID-19 positive | 360 | 1191 | 199(55%) /132(36%) /29(9%) | 60 +/- 18 |
| COVID-19 negative | 1024 | 1135 | 353(34%) /323(32%) /348(34%) | 65 +/- 19 |

* Age is given mean years +/- std.
** Numbers are after all record exclusions.

# Quantitative analysis of the model

The performance of the network was tested upon 15% (350 of 2426) of the images that were taken from the total dataset and was set aside before training. The metrics we used are accuracy, namely the proportion of successful classifications overall, sensitivity (also – recall), which is the proportion of positive images that the network classified correctly and specificity, the proportion of correctly classified negative images. We trained five deep network models whose accuracy and sensitivity rates can be seen in Table 2. We selected ResNet50 for the rest of the analysis, as it achieved the best performance in our task with accuracy 89.7% (314/350), sensitivity 87.1% (156/179) and specificity of 92.4% (158/171) on the test images. The AUC of the ROC curve is 0.95. The ROC curve is provided in Figure 2a, showing the relationship between the false positive rate (FPR) and the true positive rate (TPR) for different classification threshold values. The curve shows that for a broad range of thresholds, both a high TPR and a low FPR can be achieved. In Figure 2b we present the P-R curve, which shows the tradeoff between precision (the proportion of images labeled positive from all images that the network classified as positive) and recall as the value of the threshold is varied. This P-R curve shows a broad range of thresholds for which both high precision and high recall are attainable. The AUC of the P-R is 0.94. These ROC and P-R curves attest to the stability of the model across different confidence thresholds.

*Table 2:* **Comparison of accuracy, sensitivity and specificity of various deep networks trained and tested on the same test set.**

| Training Model | Accuracy (%) | Sensitivity (%) | Specificity (%) |
|---|---|---|---|
| ResNet34 | 89.4 (313 of 350) | 87.1 (156 of 179) | 91.8 (157 of 171) |
| **ResNet50** | **89.7 (314 of 350)** | **87.1 (156 of 179)** | **92.4 (158 of 171)** |
| ResNet50 - No preprocessing | 85.1 (298 of 350) | 82.1 (147 of 179) | 88.3 (151 of 171 |
| ResNet152 | 86.0 (304 of 350) | 83.2 (149 of 179) | 90.6 (155 of 171) |
| Chexpert | 84.0 (294 of 350) | 86.5 (155 of 179) | 81.2 (139 of 171) |
| VGG16 | 87.7 (397 of 350) | 87.1 (156 of 179) | 88.3 (151 of 171) |
| **Majority Vote** | **90.5 (317 of 350)** | **88.8 (159 of 179)** | **92.3 (158 of 171)** |

Note - The model with best accuracy and sensitivity, shown in bold, is a Majority Vote - as a vanilla (simplest) version of "ensemble" method, we gathered all algorithms' results and made a prediction by taking the label which was chosen the most.

The analysis in our paper focuses on Resnet50. **The 95% confidence intervals** for the training scores of resnet50 are: 95% CI [0.84, 0.93] for accuracy, 95% CI [0.81, 0.95] for sensitivity, and 95% CI [0.8, 0.94] for specificity. To calculate the confidence intervals, we trained the network on 10 different randomly sampled train sets, consisting of 85% of the data each, and bootstrapped each of the corresponding test sets 20 times to get a total of 200 values for each score.

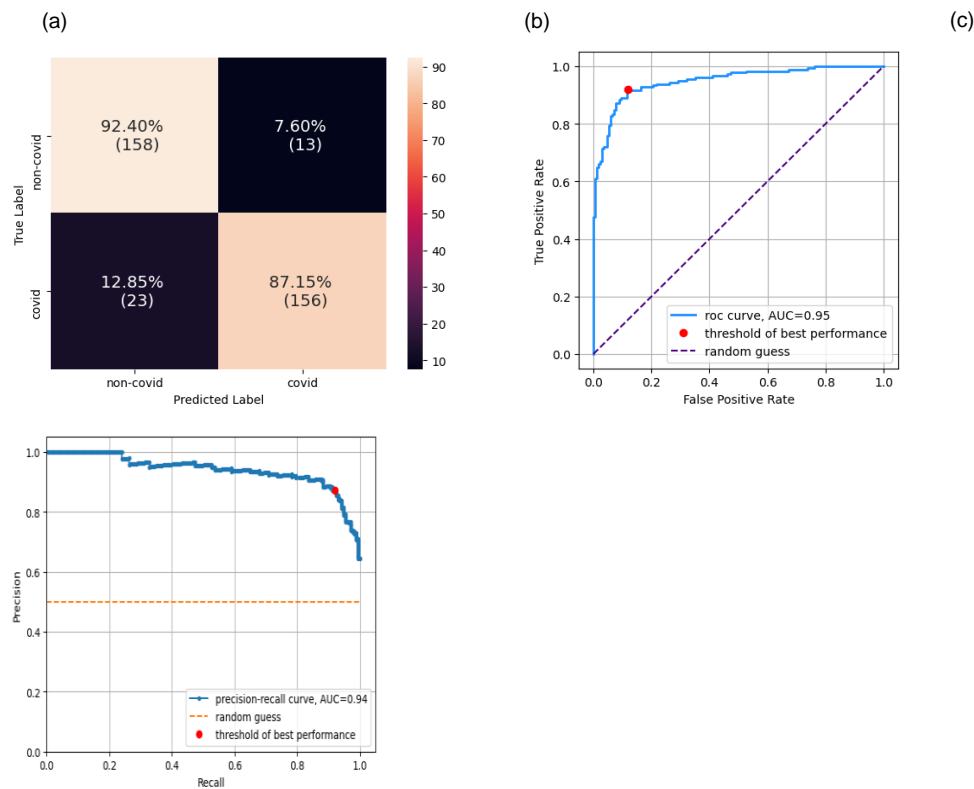

*Figure 2:* Performance of the model. **(a)** Confusion Matrix of the classification. True positive rate (TPR) at the bottom right corner, true negative rate (TNR) at the top left corner, false positive rate (FPR) at the top right corner, and false negative rate (FNR) at the bottom left corner. **(b)** Receiver Operating Characteristic (ROC) curve. The curve shows the relation between true positive rate (TPR) and false positive rate (FPR) as the threshold of the separation between positive and negative classification is varied. The performance of the model is measured by the area under the curve (AUC). Ideally, the curve should cover as much area as possible up to the upper left corner (AUC score of 1), which minimizes the FPR while maximizing the TPR. The AUC is 0.95 and a good stretch of the curve is marked on the graph where the model achieved a TPR of 87.1% and FPR of 12.8% which can be used as a threshold; **(b)** Precision-Recall curve. Shows the relation between Precision and Recall. Precision and Recall are affected from different classes of the data, thus can vary in scores when data is imbalanced (e.g. more observations of positive or negative compared to the other). We would like to have the AUC as large as possible up to the upper right corner, which maximizes both Precision and Recall. The mark on the graph represents such an optimal spot where the model achieved Precision of 92% and recall of 87%.

The 95% Confidence interval (CI) for AUC-ROC score is [0.91, 0.97]. The 95% CI for AUC-Precision recall curve is [0.91, 0.97].

We then train ResNet50 on the dataset with and without all the preprocessing stages. As seen in

Table 2, preprocessing incurs an improvement of 4% in accuracy and 5% in sensitivity.

## Qualitative analysis of the model

In addition to the binary decision of whether a patient has COVID-19, we provide a score between 0 and 1, corresponding to the probability the network assigns to the positive label. It is given by the activation of the network's last layer, before it is passed through an activation function that produces the binary output. Whenever this score is above the threshold of 0.5, an image is classified as positive for COVID-19. We generate a histogram of these scores, as can be seen in Figure 3, and observe that the majority of the correctly classified points are accumulated at the edges, while the wrongly classified images are more spread out along the x-axis.

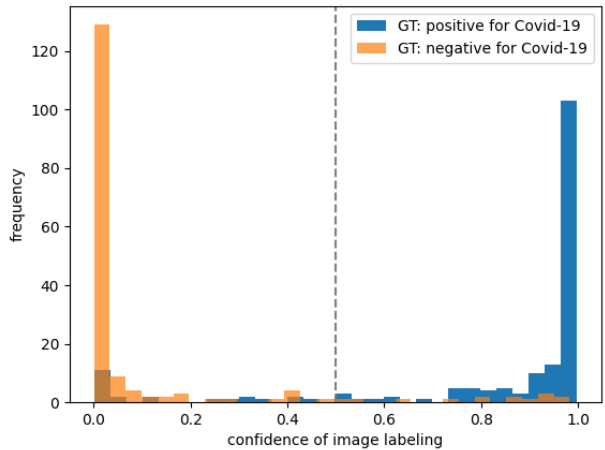

*Figure 3:* Classification score histogram. Ground truth (GT) labels are in colors. Every image is scored on a scale between 0 and 1 with threshold of 0.5, seen as a dashed line, such that all images with a higher score will be classified as positive for COVID-19 and images below as negative. Negatively labeled images that received a score above 0.5 are, therefore, incorrectly classified images, and vice versa with respect to positively labeled images. However, the closer the image score is to one of the edges (0 or 1), the stronger the confidence in the image's classification. The accumulation of two distinct colors on the edges point to good separation of many observations with strong confidence in the classification.

We additionally visualize the distinction made by the model using t-distributed Stochastic Neighbor Embedding (t-SNE) (18). t-SNE uses a nonlinear method to reduce high dimensional vectors into two dimensions, making it possible to visualize the data points and reveal similarities and dissimilarities between them. We used one of the last layers of the networks, which essentially provides an embedding of the images into a vector space. These vector embeddings of the images are given as input to the t-SNE. In Figure 4 it can be seen that the arrangement of the dots, representing the features of the images, colored by their GT labels. The figure depicts two distinct clusters, revealing a similarity between most of the images with the same GT.

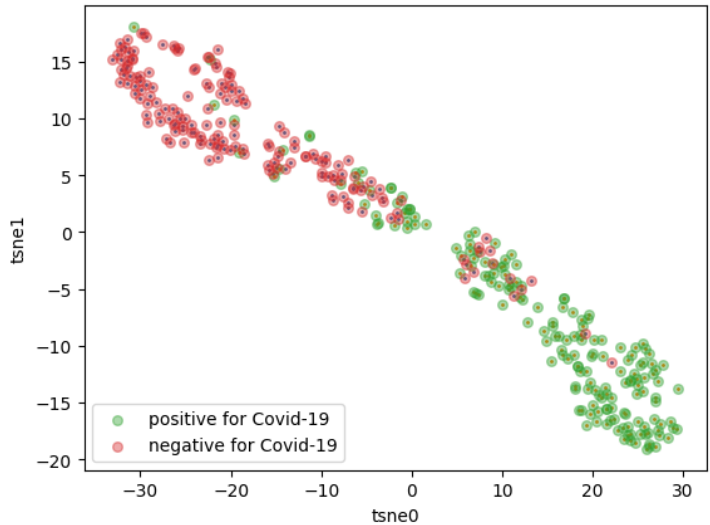

*Figure 4*: t-distributed Stochastic Neighbor Embedding (t-SNE). A high- dimensional feature vector is extracted for each image from one of the decision layers, which are used for decision of the output of the neural network, and is reduced into 2 dimensions. Each point on the graph represents the features of an image after dimension reduction and arrangement in space. Next the images were colored according to their ground truth (GT), thus revealing two main clusters. The clusters are mostly in one color each, which essentially shows a strong association of the features, extracted from the decision layer and are used to arrange in space, with the GT of the images, represented by the colors.

In order to test the model on a more difficult task, we were supplied with 22 CXRs, 9 positive for COVID-19 and 13 as control, classified by radiologists as difficult to diagnose and used as a test on our model. The accuracy on the test was 77% and sensitivity of 77%. In Figure 5, three correctly classified images from this test are shown with the network's classification score and the GT.

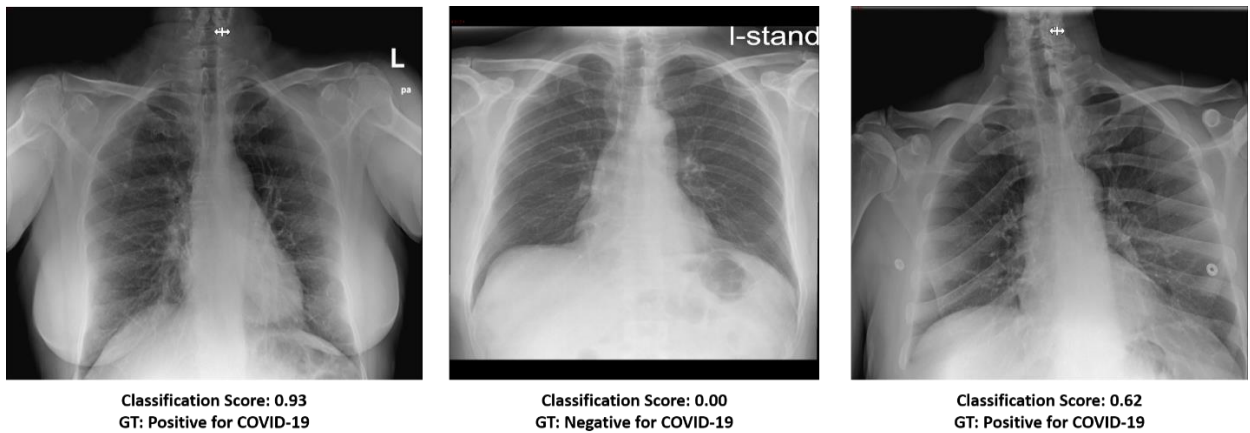

*Figure 5:* Three images labeled by a radiologist as hard to diagnose. Despite this, the model was able to classify

them correctly. Each image is scored with classification score on a scale between 0 and 1 with threshold of 0.5 such that all images with confidence score above the threshold will be labeled as positive for COVID-19 and images below as negative. The ground truth (GT) of each image is also shown.

Finally, we applied K-Nearest Neighbors (KNN) on the image embeddings in order to retrieve images similar to each other as shown in Figure 6. For each image we retrieve 4 images with the closest image embeddings; averaging over these images' predictions achieves 87% accuracy (305/350) and 83.2% sensitivity (149/179), meaning that the nearest images typically have the same labels.

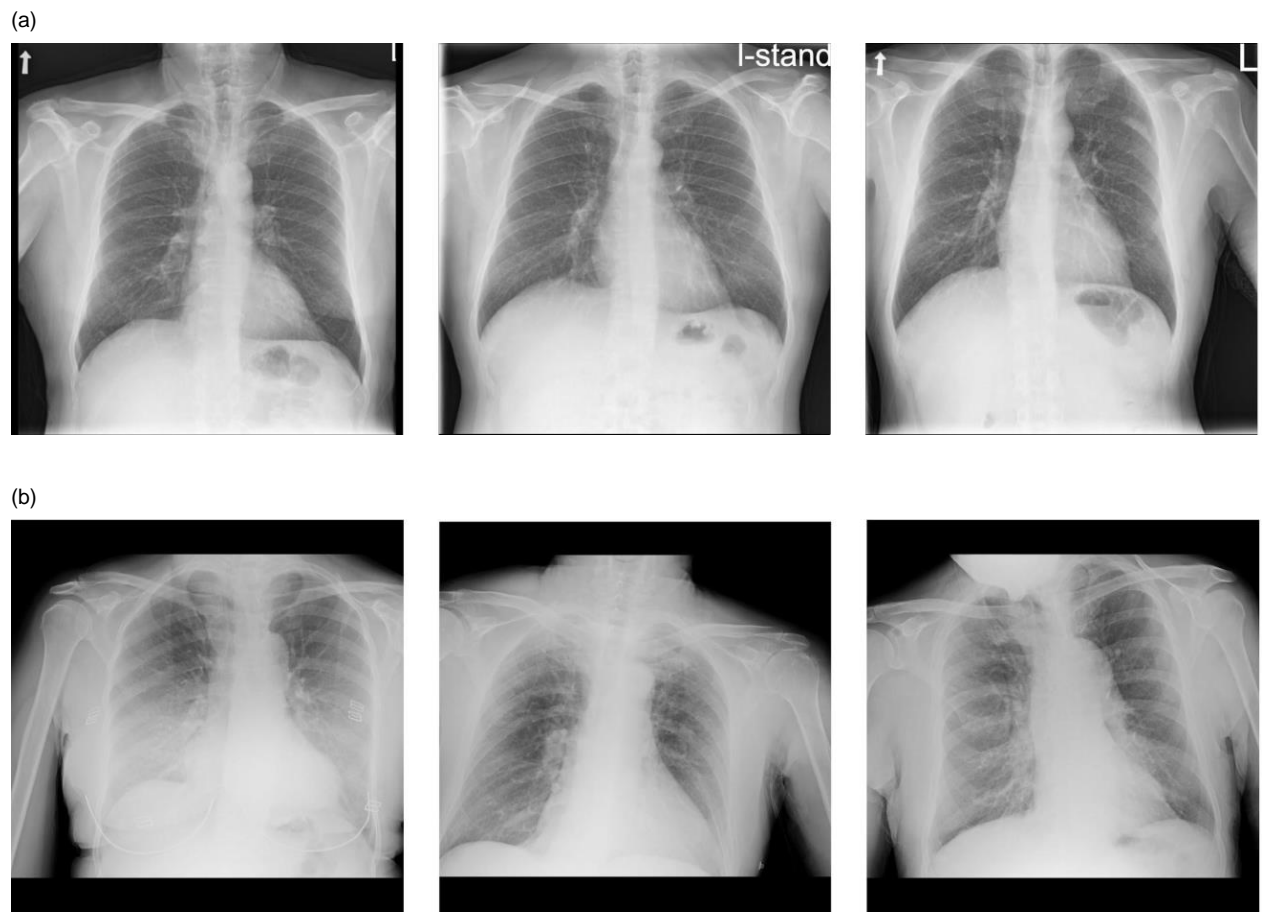

Figure 6: **In the figures,** the left image is a CXR from the test set, and the two on the right are the two images closest to it from the training set, given the image embeddings from the network's last layer. **(a)** All three images are **COVID-19 Negative.** The distances between the middle and rightmost images to the left one are 0.54 and 0.56 respectively. **(b)** All three images are **COVID-19 positive.** The distances between the middle and rightmost images to the left one are 0.51 and 0.55 respectively. The overall mean distance between training and test images is 3.9+/-2.5 (mean +/- std). The mean distance between all positive training and positive test images is 1.4+/-1.9, between negative training and negative test images 2.2+/-1.3, and between images from different

classes is 5.8 +/-1.9. Images from different classes are further away from each other, but whether a close distance truly corresponds to similar lung findings still requires verification.

# 4 Discussion

In this study we developed a deep neural network pipeline to classify chest X-ray (CXR) images of patients as coronavirus disease 2019 (COVID-19) positive or negative, and to identify which X-ray scans are similar to each other. The dataset we used is compiled to be as representative as possible of images from patients that would enter a healthcare unit with a suspicion of COVID-19 in a real clinical scenario. The ResNet50 model we trained for classification achieved 89.7% accuracy, 87.1% sensitivity and area under Receiver Operating Characteristic (ROC) curve of 0.95 in classifying COVID-19 images. In addition, we created a tool that retrieves the CXR images most similar to a given image. This can provide physicians with a reference to previous patients that had similar lung findings. They can use the internal information they have from the hospital about these previous patients to infer decisions upon treatment.

The method for image retrieval has not previously been well established in the literature in the context of COVID-19 CXR. Other groups have worked on COVID-19 classification using neural networks, mostly based on publicly available image sources, such as COVID-19 image data collection (19) with 481 COVID-19 positive images and COVID-Net open source (20) initiative with 473 COVID-19 X-ray images. The performance reported in these research papers is generally high with accuracy rate ranging from 89%-99% and specificity ranging from 80% to 100% (19,21).

However, these results were obtained via testing solely on subsets of the data available to the research. These have a number of drawbacks. They include a limited number of positive COVID-19 CXR images, which can cause the model to overfit, as it is exposed to a relatively small number of characteristics from the data which can impair the ability to generalize to external datasets. These models' reliability still need to be verified on external data. As machine learning models tend to improve and generalize better when the amount of data increases (22), a dataset with more positive COVID-19 images as the one used in this study, with 1191 positive CXR, tends to be more stable. In addition, these datasets were compiled from various sources, often using one source only for COVID-19 images and another only for COVID-19 negative images. Positive and negative images in these datasets may therefore be produced by different X-ray machines, in particular portable and fixed machines, which give rise to images with different expressions of optical features. This can allow the network's predictions to rely on features related to the source more than on the relevant medical information (23). In this research we used CXR from the same machines both for patients with both positive and negative COVID-19 outcomes.

As future work, we intend to deploy our model for testing in a clinical setting. We will further investigate the scoring process for the image similarities we provide. We would ideally like to compare the disease progression for patients that were found by our tool to have similar lung findings. Additionally, we will examine how CXR are influenced by progression of the disease. Lung damage may remain after the virus leaves the body, leading to false positives in classification in later stages of the disease. Lastly, our classifier is tailored towards portable X-rays within the four Israeli hospitals that provided the data. It may need further fine tuning to be used in other hospitals or diagnostic settings.

In summary, we showed a deep neural network which is able to reliably detect patients with coronavirus disease 2019. Even though medical imaging has not yet been approved as a standalone diagnosis tool (9), we believe it can be used as an aid to medical judgement with the advantage of immediate outcome. We also created a tool for X-ray image retrieval based on lung similarities. This tool can help physicians draw connections between patients with similar disease manifestations, by referring them to images with similar lung characteristics. These images can be linked internally to the corresponding patients, and the treatment and outcome of these patients can then inform their decision upon treatment for the current patient.

## Acknowledgements

We would like to acknowledge Avithal Elias and Nadav Nehmadi for their helpful comments and contribution in the initial stages of the project.

# Appendix

In this appendix, we elaborate further on the data processing and the neural network design.

## 1 Data preprocessing

Before training, each image goes through a preprocessing pipeline. We start by cropping out areas that contain only text around the images themselves. We then unify the image sizes, preserving the original aspect ratios via padding, and apply a CLAHE (filter that was seen to enhance images and improve deep learning performance[10]). On the training data, we also apply a series of augmentations.

## Augmentation

Augmentations are transformations performed on the data that serve a dual purpose. First, applying the augmentations creates additional diverse set of images from the existing ones and enables one to artificially increase a dataset to improve performance[11]. Augmentations are therefore very commonly used on medical images, where datasets tend to be relatively small[12]. Second, these transformations can help the network generalize better[13], as they alter features that are unimportant to the identification of COVID-19 in the lungs. This way the network can learn the important features and ignore the irrelevant ones. Crucially, the transformations must preserve the image labels - a coronavirus patient must still be identifiable as one. To ensure this, we consulted with radiologists when defining the transformations and their parameter ranges. The augmentations are performed randomly, with parameters chosen uniformly within the defined range as seen in Figure 1. Not all augmentations are applied each time, but rather each augmentation has a certain probability of being applied, represented by p below:

---

[10] "Classification of Breast Microscopic Imaging using Hybrid ...." https://ieeexplore.ieee.org/document/8844937/. Accessed 23 Aug. 2020.
[11] "The Effectiveness of Data Augmentation in Image ...." 13 Dec. 2017, https://arxiv.org/abs/1712.04621. Accessed 23 Aug. 2020.
[12] "Data Augmentation in Training Deep Learning Models for ...." 16 May. 2020, https://link.springer.com/chapter/10.1007/978-3-030-42750-4_6. Accessed 23 Aug. 2020.
[13] "Data Augmentation in Training Deep Learning Models for ...." 16 May. 2020, https://link.springer.com/chapter/10.1007/978-3-030-42750-4_6. Accessed 23 Aug. 2020.

1. brighten, p=0.4

2. gamma contrast, p=0.3

3. CLAHE, p=0.4

4. rotate d ∈ [7,7] degrees p=0.4

5. shear d ∈ [7,7] degrees p=0.4

6. scale up to 0.2 on each axis p=0.4

7. flip from left to right, p=0.5

8. either sharpen or apply Gaussian blur

9. horizontal flip, p =0.5

We decided to apply left to right flips, as COVID-19 is known to affect the lungs symmetrically. Thus, flipping will not change the characteristic manifestation of the disease. Moreover, some X-ray images may be taken from the back, and we do not always have clear labels as to the direction in which the X-ray was taken. Adding flips of the images can make the network robust to this.

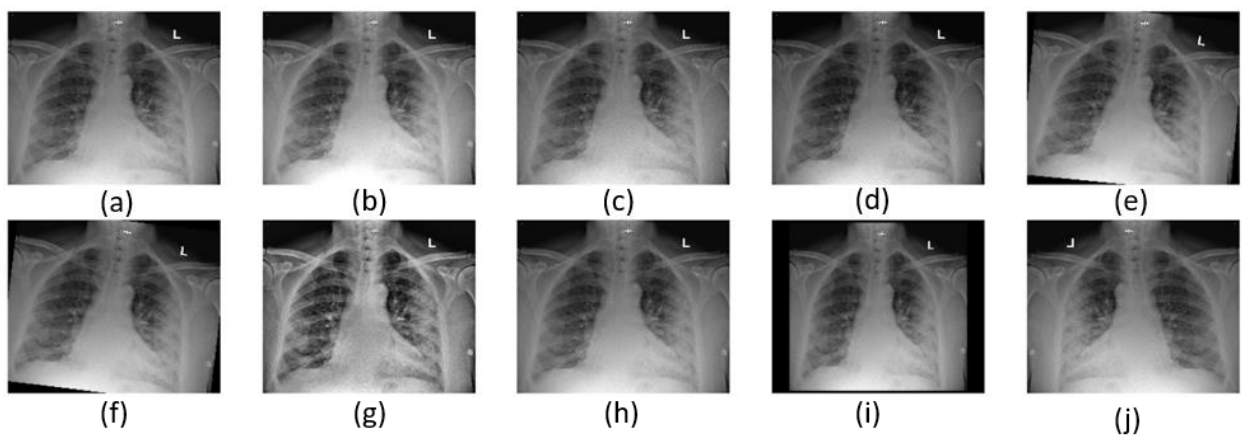

*Figure 1:* Image augmentation. In order to increase the number of images which can improve training performance, several different transformations are performed with a certain probability. The transformations showed: On top: (a) Original image, (b) Brighten, (c) Sharpen, (d) Gamma contrast, (e) Shear. On bottom: (f) Rotate 7 degrees, (g) CLAHE, (h) Gaussian blur, (i) Scale, (j) Horizontal flip.

# Segmentation

A novel aspect of our model architecture is adding an additional input channel to each image in the form of a probability vector, which indicates for each pixel the probability it belongs to the lung. These probabilities are obtained by applying a pre-trained U-net to segment the lung area from the image. Adding this mask as an additional channel to the X-ray image helps the network focus on the lung area while training. An example of segmentation can be seen in Figure 2.

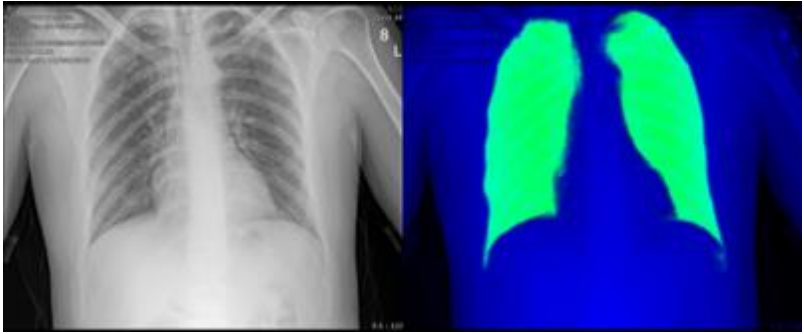

*Figure 2:* **Lung segmentation using a U-net architecture**

# 2 Network architecture

Deep learning-based automated diagnosis approaches have been gaining interest in recent years, mainly due to their ability to extract sophisticated features from images. This allows to describe an image in an alternative way from which we can derive computational conclusions. Based on that, our network architecture consists of two main parts - feature extraction, and decision head. The feature extractor is a neural network based on a Resnet50 architecture that gets an image as input (in our case - 2D image), performs mathematical operations on it and outputs a feature map, namely a matrix of numbers which describe the image. This matrix of features is converted to a vector (with the same values) and then goes into the decision head which is a simple neural network. In our case it consists of 3 fully connected layers. The output of the decision head is two numbers which describe the confidence of the algorithm about the classification results: COVID-positive or COVID-negative. In addition, the last layer (a vector) in the decision head is referred to as the "embedding" and is used as an input to the t-SNE and KNN algorithms described in the text.

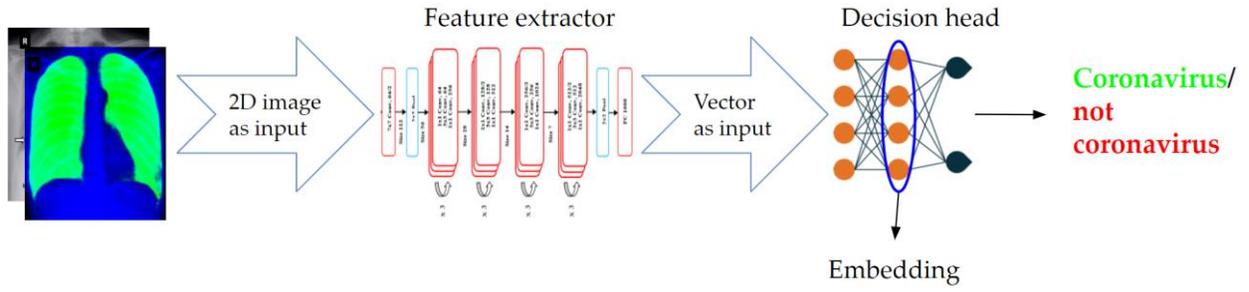

*Figure 2:* Pipeline of the neural network stage in inference time. Input images are passed through a sequence of convolutional layers that extract lower-dimensional vector representations for each image; these representations are optimized for the task at hand, in our case - separation in the vector space between images belonging to different label classes.